\documentclass[%
 reprint,
 amsmath,amssymb,
 aps,
]{revtex4-1}

\usepackage{graphicx}
\usepackage{dcolumn}
\usepackage{bm}

\usepackage{chemformula, multirow, filecontents}
\newcommand*{\affaddr}[1]{#1} 
\newcommand*{\affmark}[1][*]{\textsuperscript{#1}}

\begin{document}

\preprint{APS/123-QED}

\title{The role of the interface in controlling the epitaxial relationship between orthorhombic $\text{LaInO}_\text{3}$ and cubic $\text{BaSnO}_\text{3}$}

\author{%
Martina Zupancic\affmark[1], Wahib Aggoune\affmark[2], Toni Markurt\affmark[1], Youjung Kim\affmark[3], Young Mo Kim\affmark[3], Kookrin Char\affmark[3], Claudia Draxl\affmark[2], and Martin Albrecht\affmark[1]\\
\affaddr{\affmark[1]\textit{Leibniz-Institut f\"ur Kristallz\"uchtung, Max-Born-Str. 2, 12489 Berlin, Germany}}\\
\affaddr{\affmark[2]\textit{Institut f\"ur Physik and IRIS Adlershof,  Humboldt-Universit\"at zu Berlin, 12489 Berlin, Germany}}\\
\affaddr{\affmark[3]\textit{Institute of Applied Physics, Dept. of Physics and Astronomy, Seoul National University, Seoul 08826, Korea}}
}

\begin{abstract}
Epitaxial perovskite oxide interfaces with different symmetry of the epitaxial layers have attracted considerable attention due to the emergence of novel behaviors and phenomena.  In this paper, we show by aberration corrected transmission electron microscopy that orthorhombic \ch{LaInO3} films grow in form of three different types of domains on the cubic \ch{BaSnO3} pseudosubstrate. Quantitative evaluation of our TEM data shows that $c_{pc}$-oriented and $a_{pc}/b_{pc}$-oriented domains are present with similar probability. While continuum elasticity theory suggests that $c_{pc}$-oriented domains should exhibit a significantly higher strain energy density than $a_{pc}/b_{pc}$-oriented domains, density functional calculations confirm that $c_{pc}$- and $a_{pc}$-oriented domains on \ch{BaSnO3} have similar energies.

\end{abstract}

\maketitle

\section{\label{sec:Introduction}Introduction}

Over the last decade, many studies have focused on polar - non-polar perovskite oxide interfaces due to the promise to realize novel electronic devices, and to tune the electronic behavior between metallic, semiconducting, and superconducting \cite{ohtomo2004high, yeh1999nonequilibrium, bjaalie2014oxide}. The formation of a 2-dimensional electron gas at the interface, as well as the mobility of carriers and other physical properties, crucially depend on the atomic structure of the interface. Since most of these heterostructures are heteromorphous and consist of materials with different space groups, the crystallographic orientation relationship between them is an important issue. A common approach to predict the preferred orientation in coherently strained films is to consider the orientation that results in the lowest total energy, $E_{total}$, where: 
\begin{equation}
\label{eq:energy}
E_{total}=E_{interface}+E_{strain}.
\end{equation}

In classical group IV semiconductors or III-V compounds, the interface energy in most cases is negligible, and the total energy is dominated by the strain energy. The strain induced by the substrate in these systems is accommodated by a tetragonal distortion of the unit cell and can be calculated by continuum elasticity theory. Epitaxial growth of \ch{ABO3} perovskite oxides on each other is governed by a number of peculiarities:

\begin{itemize}
\item Perovskites in most cases exhibit different crystal symmetry (e.g. perfect cubic, rhombohedral, orthorhombic) mediated by symmetry reducing distortions of the perfect cubic structure, often reflected in a respective tilt of the \ch{BO6} octahedra. If, for example, a layer with lower symmetry grows coherently on a substrate with perfect cubic symmetry, different epitaxial relationships are possible. 
\item Perovskites can adopt strain by both, octahedral rotations (along the surface normal or perpendicular to it) and by relative displacement of cations and anions, i.e. by distortions of the unit cell. Peculiarities of the strain accommodation will strongly depend on the chemistry of the compound \cite{rondinelli2011structure}. 
\item The interface may play a role in the energetics of the system. The octahedra tilt pattern of the substrate may influence the one of the epitaxial layers and counteract strain accommodation. Chemistry may then control the energy balance over strain. Another important contribution that controls interface formation is the charge that is present at a polar - non-polar interface.  
\end{itemize}

In this paper, we combine quantitative high-resolution transmission electron microscopy (HRTEM) and ab-initio density functional theory (DFT) calculations to study the formation of the polar - non-polar interface between the orthorhombic perovskite \ch{LaInO3} and the cubic perovskite \ch{BaSnO3}. This system has attracted considerable attention in the last decade, since it enables the formation of a 2-dimensional electron gas at its interface, similar to the prototypic system \ch{LaAlO3} on \ch{SrTiO3} \cite{kim2016conducting, kim2018laino3, kim2019interface, ohtomo2004high}. In addition, \ch{BaSnO3} possesses the highest electron mobility ($\sim 300$ cm\textsuperscript{2}/Vs) among the transparent conductive oxides \cite{kim2012physical, kim2012high}.

Here we show that orthorhombic \ch{LaInO3} grows coherently in domains with three possible orientations on the cubic \ch{BaSnO3} pseudosubstrate. Despite significant difference in strain energy, all domains are present with similar probabilities. Since the perovskite oxides often exhibit phase transformations in the range between room temperature and typical film growth temperatures, it is not clear whether the domains form already during growth or just at a later stage of cool-down after growth. The presence of strain may also shift the transition temperatures, therefore in this work we perform TEM in-situ heating experiment to check if \ch{LaInO3} undergoes phase transitions \cite{inaba2001structural, choi2010phase, schlom2007strain}.

While similar observations of differently oriented domains have been made for the growth of orthorhombic \ch{SrRuO3} and \ch{CaRuO3} films on cubic substrates \cite{jiang1998domain, proffit2008influence}, a consisted explanation is not given. We show that the tilt pattern of the growing layer is controlled by the cubic substrate and causes the shifts of the total energy such that different orientations are energetically degenerated.

\section{\label{sec:Methods}Methods}

\ch{BaSnO3}/\ch{LaInO3} heterostructures studied in this paper were grown by Pulsed Laser Deposition (PLD) on \ch{TiO2}-terminated \ch{SrTiO3} (001) substrates at 750 $^{\circ}$C in $100$ mTorr of oxygen pressure using KrF excimer laser with energy fluence in the range of $1.2\sim 1.5$ J/cm\textsuperscript{2}.  All targets were provided by Toshima Manufacturing Co. in Japan.

High-resolution TEM was performed with an aberration corrected FEI Titan 80-300 operated at 300 kV, with the corrector for spherical
aberrations (Cs) set to a Cs = -15 $\mu$m. Cross-sectional TEM samples were prepared along the $\langle100\rangle$ lattice direction of the \ch{BaSnO3} pseudosubstrate by tripod polishing and argon ion-milling at liquid nitrogen temperature. Ar\textsuperscript{+} ion-milling was done by a precision ion polishing system (PIPS) at beam energies from 4.0 to 0.2 keV. Plane-view samples were prepared by wedge polishing with a wedge angle of 4$^{\circ}$, using the Allied MultiPrep\textsuperscript{TM} system.
 
The lamella used for the in-situ heating TEM experiment was prepared by cutting a piece of the wedge-polished sample by focused ion beam (FIB) and transferring it by in-situ lift-out method on a Protochips' Fusion E-chip. In-situ heating experiment was conducted with a Protochips' Fusion holder. The sample was heated in vacuum with a 5 $^{\circ}$C/s ramp rate in the temperature range from 25–750 $^{\circ}$C.

Ground-state properties are calculated using DFT, within the generalized gradient approximation (GGA) for the exchange-correlation functional in the PBEsol parameterization \cite{perdew2008restoring}. All calculations are performed using FHI-aims \cite{blum2009ab}, an all-electron full-potential package. The code is based on numerical atom-centered orbitals. For all atomic species we use $tight$ settings with the $tier$ 2 basis set for oxygen (O), $tier1+fg$ for barium (Ba), $tier1+gpfd$ for tin (Sn), $tier1+hfdg$ for lanthanum (La), and $tier1+gpfhf$ for indium (In). The self-consistent field convergence criteria are $10^{-6}$ electrons for the density, $10^{-6}$ eV for the total energy, $10^{-4}$/{\AA} for the forces, and $10^{-4}$ eV for the eigenvalues. For bulk \ch{BaSnO3} and \ch{LaInO3}, both, lattice constants and internal coordinates are optimized until the residual forces on each atom are less than 0.001 eV/{\AA}. The sampling of the Brillouin zone is performed with an $8\times 8 \times 8$ k-grid for bulk \ch{BaSnO3}, and with an $6\times 6\times 4$ k-grid for bulk \ch{LaInO3}. These parameters ensure a numerical precision better than 5 meV/atom for the total energy and 0.001 {\AA} for the lattice parameters. For the \ch{BaSnO3}/\ch{LaInO3} heterostructure, the Brillouin zone sampling is performed with an $6\times 6\times 1$ k-grid for the case where the 
$c_{pc}$ orientation of \ch{LaInO3} is parallel to the
\ch{BaSnO3} (001) surface, and an $4\times 4\times 1$ k-grid when the $a_{pc}$ orientation of \ch{LaInO3} is parallel to the \ch{BaSnO3} (001) surface. We include a vacuum of $\sim 70$ {\AA} and apply a dipole correction in the non-periodic [001] direction, in order to prevent unphysical interactions between neighboring replica. In this case, only internal coordinates are optimized until the residual forces on each atom are less than 0.001 eV/\AA. The lattice parameter of the first unit cell of the \ch{BaSnO3} pseudosubstrate is fixed to the bulk value. To calculate the strain energy density at the \ch{BaSnO3}/\ch{LaInO3} interfaces, we compute the elastic constants of bulk \ch{BaSnO3} and \ch{LaInO3} from the second derivatives of the total energy \cite{golesorkhtabar2013elastic}. The $ElaStic$ package is used to compute the second derivatives and extract the elastic constants \cite{golesorkhtabar2013elastic}. Atomic structures are visualized using the VESTA software \cite{momma2011vesta}.

\section{\label{sec:Crys}Crystallographic considerations and basic materials parameters}

\begin{figure}[b!]
\includegraphics[width=0.47\textwidth]{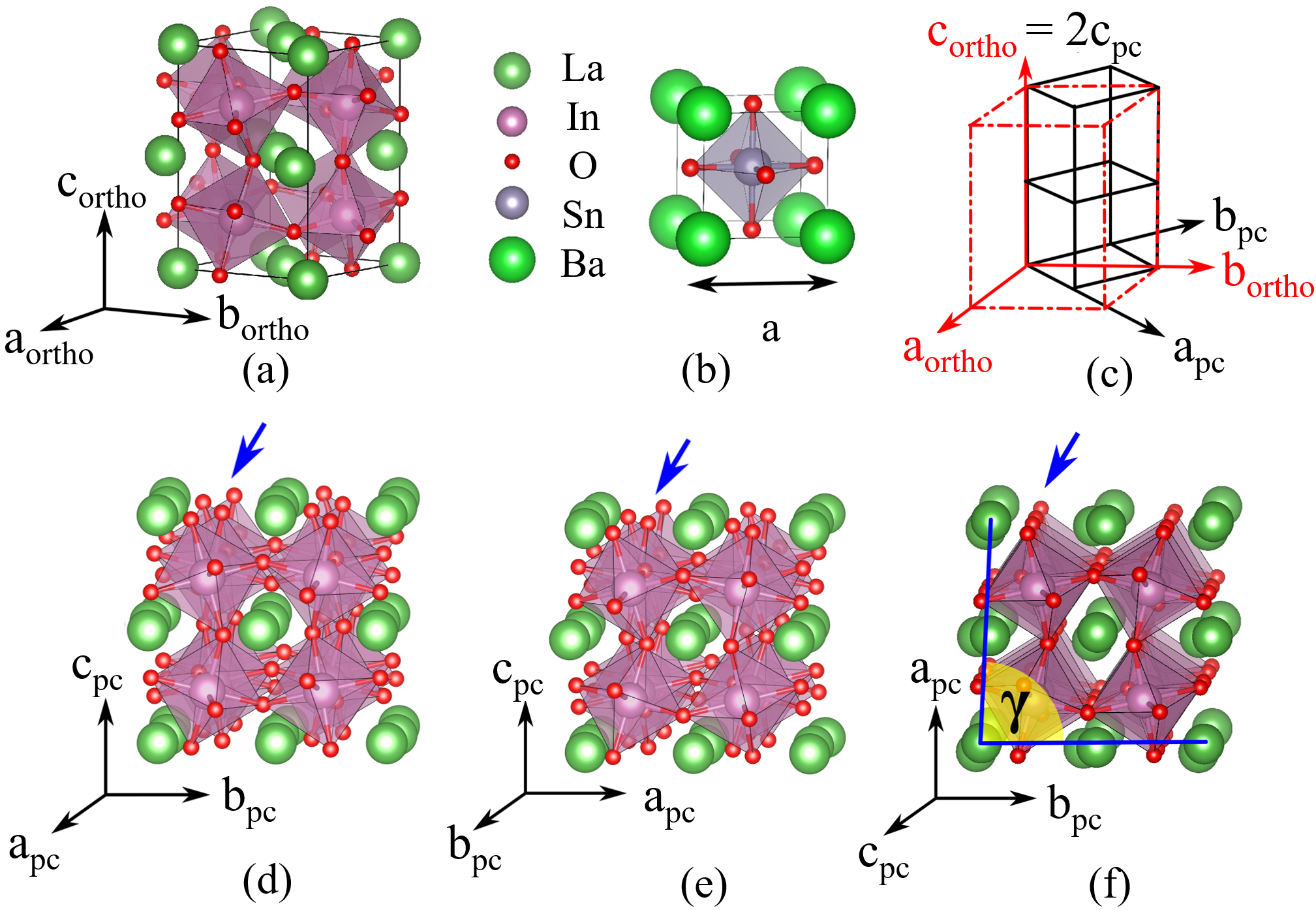}
\caption{\label{fig:Fig1}Sketch of the primitive unit cell of (a) orthorhombic \ch{LaInO3} and (b) cubic \ch{BaSnO3}. (c) Scheme of the relationship between the  orthorhombic (red) and pseudocubic (black) \ch{LaInO3} unit cell. The a\textsuperscript{-}a\textsuperscript{-}c\textsuperscript{+} Glazer tilt pattern describes \ch{InO6} octahedral tilt (marked with blue arrows) in orthorhombic \ch{LaInO3} perovskite structure: (d) a\textsuperscript{-} out-of-phase tilt around the $a_{pc}$ = $[100]_{pc}$ axis (out-of-phase tilt signifies oxygen octahedra which rotate along the pseudocubic rotational axis in opposite direction), (e) a\textsuperscript{-} out-of-phase tilt around the $b_{pc}$ = $[010]_{pc}$ axis and (f) c\textsuperscript{+} in-phase tilt around the $c_{pc}$ = $[001]_{pc}$ axis (in-phase tilt signifies oxygen octahedra which rotate along the rotational axis in the same direction). The angle between $a_{pc}$ and $b_{pc}$, $\gamma$, equals 87.6$^{\circ}$.}
\end{figure}

\ch{BaSnO3} is a perfect cubic perovskite with a lattice constant of 4.116 {\AA} (Fig. \ref{fig:Fig1} (b)) \cite{maekawa2006thermal}. \ch{LaInO3} has an orthorhombic structure of type \ch{GdFeO3} (Fig. \ref{fig:Fig1} (a)) with lattice parameters $a_{ortho}$ = 5.9404 {\AA}, $b_{ortho}$ = 5.7229 {\AA}, and $c_{ortho}$ = 8.2158 {\AA}, \cite{park2003lanthanum}. The structure of  \ch{LaInO3} is characterized by rotation of the \ch{InO6} octahedra along all three pseudocubic directions and defined by the a\textsuperscript{-}a\textsuperscript{-}c\textsuperscript{+} tilt pattern according to the Glazer notation \cite{glazer1972classification}, which  is described in Figs.  \ref{fig:Fig1} (d), (e), and (f). A scheme of the \ch{LaInO3} orthorhombic unit cell and its correlation with the pseudocubic one is shown in Fig. \ref{fig:Fig1} (c), where $a_{pc}$, $b_{pc}$, and $c_{pc}$ represent [100], [010], and [001] pseudocubic directions, respectively. When comparing the $a_{pc}$ and $b_{pc}$ film growth orientations, there is no difference with respect to the strain between them. These two orientations are the same if we compare an area larger than one unit cell. However, if we consider just one unit cell (Figs. \ref{fig:Fig1} (d) and (e)), a different rotation direction around the in-plane axis for the equivalent oxygen atoms in the oxygen octahedra can be observed. 

\begin{table}[h!!!]
\caption{\label{tab:latticeParameters}Experimental lattice parameters (in \AA) of \ch{BaSnO3} and \ch{LaInO3} compared to theoretical values obtained by PBEsol (this work).}
\centering
\footnotesize
\begin{tabular}{c c c c c}
\hline \hline
 & & $a$ & $b$ & $c$ \\ \hline
\multirow{3}{*}{\ch{BaSnO3}} & \multirow{2}{*}{Expt.} & 4.117 \cite{smith1960some, mizoguchi2004probing} & & \\
 & & 4.116 \cite{maekawa2006thermal}& - & - \\ 
 & This work & 4.119 & & \\ \hline
 \multirow{2}{*}{\ch{LaInO3}} & Expt. \cite{park2003lanthanum} & 5.940 & 5.723 & 8.216\\
 & This work & 5.939 & 5.698 & 8.210 \\ \hline
\hline
\end{tabular}
\end{table}

Table \ref{tab:latticeParameters} compares experimental lattice parameters to those calculated in this work. The computed lattice parameters of both \ch{BaSnO3} and \ch{LaInO3} are in good agreement with previously reported experimental values \cite{maekawa2006thermal, park2003lanthanum, smith1960some, mizoguchi2004probing}, and are used in the following calculations of the strain energy density based on elastic continuum theory. In Table \ref{tab:elasticConstants}, we summarize the calculated elastic constants of orthorhombic \ch{LaInO3} which are used in the following calculations. Our results are in good agreement with previous theoretical findings based on DFT calculations reported in \cite{bouhemadou2010structural, erkicsi2016first}. Small differences in the elastic constants of \ch{LaInO3} are mainly due to the differences in the lattice parameters.

\begin{table*}[!htb]
\caption{\label{tab:elasticConstants}Calculated elastic constants of \ch{LaInO3} and \ch{BaSnO3} in GPa.}
\centering
\setlength{\tabcolsep}{10pt}
\begin{tabular}{c c c c c c c c c c c}
\hline \hline
 & & $C_{11}$ & $C_{12}$ & $C_{22}$ & $C_{13}$ & $C_{23}$ & $C_{33}$ & $C_{44}$ & $C_{55}$ & $C_{66}$ \\ \hline
\multirow{2}{*}{\ch{BaSnO3}} & Ref. \cite{bouhemadou2010structural} & 285.2 & 68.5 & & & & & 84.3 & & \\
 & This work & 286.5 & 83.4 & & & & & 93.6 & & \\ \hline
 \multirow{2}{*}{\ch{LaInO3}} & Ref. \cite{erkicsi2016first} & 238.1 & 121.7 & 225.8 & 111.5 & 104.4 & 195.3 & 52.6 & 70.4 & 52.6\\
 & This work & 243.9 & 129.1 & 234.8 & 118.8 & 112.2 & 204.7 & 54.7 & 73.6 & 58.1\\ \hline \hline

\end{tabular}
\end{table*}

\section{\label{sec:Exp}Experimental results}

\subsection{\label{sec:domains}Domain distribution}

\begin{figure}
\includegraphics[width=0.47\textwidth]{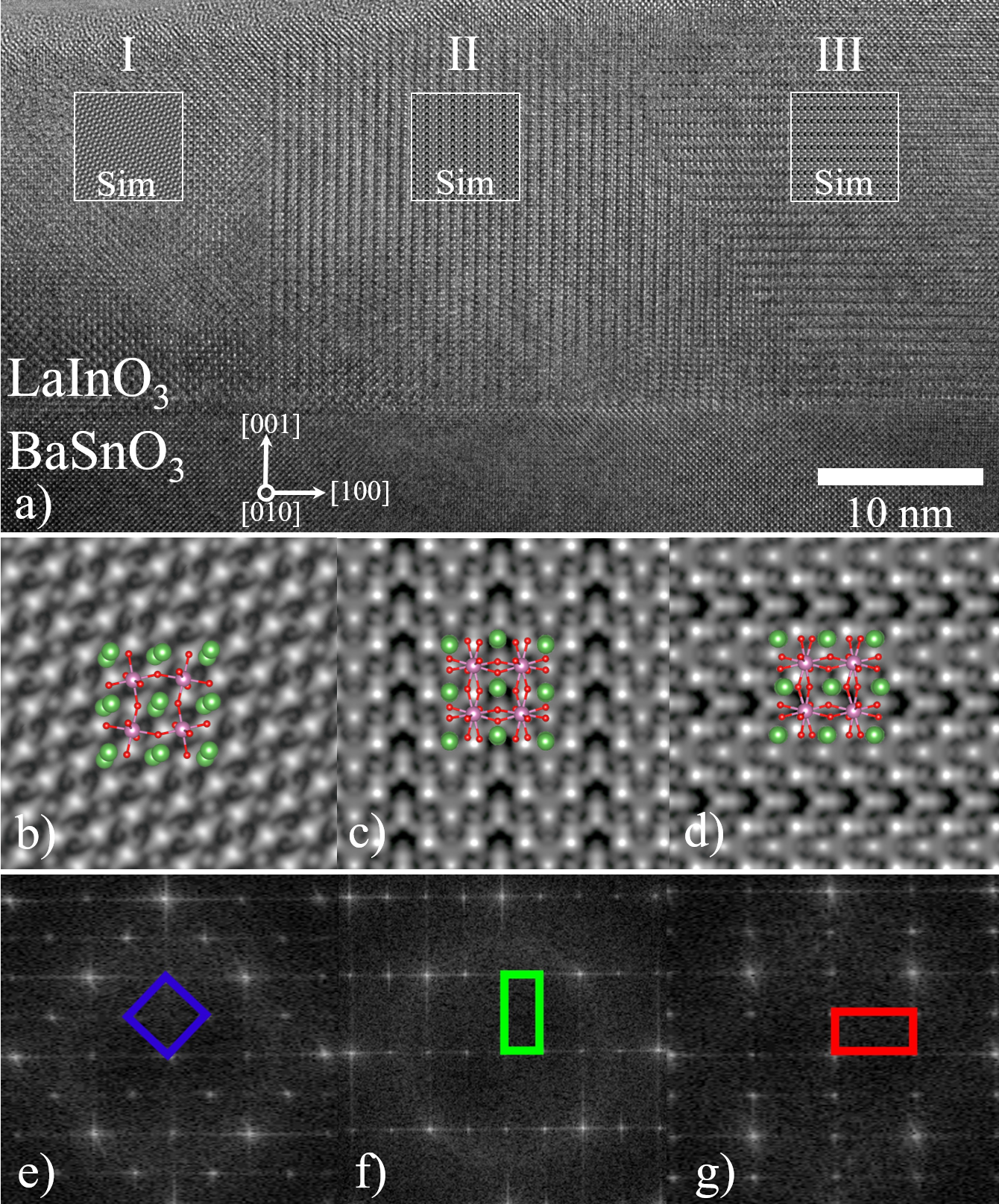}
\caption{\label{fig:Fig2}(a) Cross-sectional HRTEM image of the \ch{LaInO3} film grown on \ch{BaSnO3}. HRTEM simulations are shown as insets. Corresponding magnified simulations are shown together with assigned atomic models in (b), (c), and (d). Fast Fourier transformed images of \ch{LaInO3} domains are shown on (e), (f), and (g) where blue, green, and red marked unit cells correspond to the domains I, II, and III shown in (a), respectively.}
\end{figure}

\begin{figure}[!htb]
\includegraphics[width=0.47\textwidth]{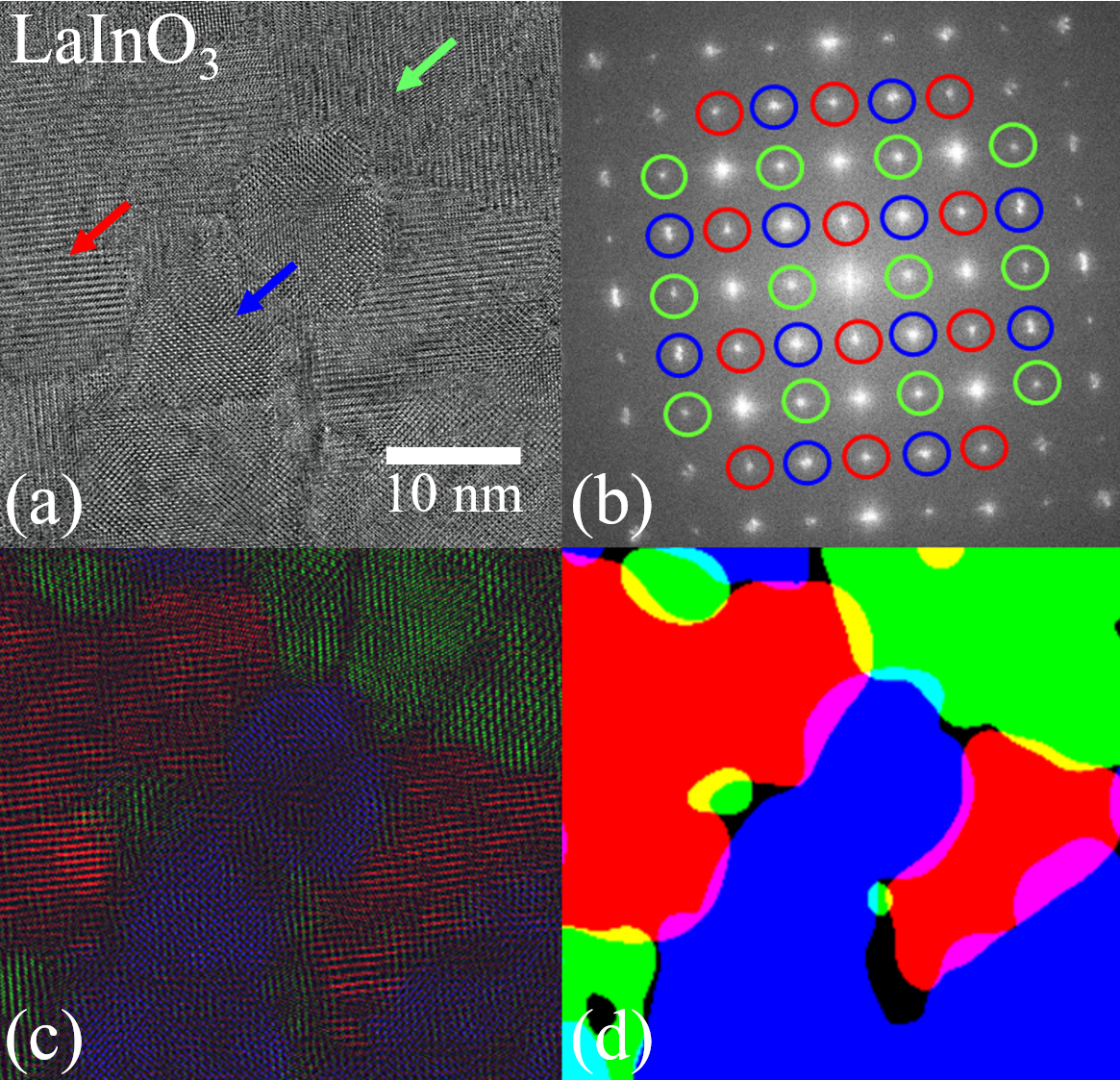}
\caption{\label{fig:Fig3}(a) HRTEM image of plane-view. (b) Fast Fourier transformation of HRTEM image where red, green, and blue correspond to reflexes specific only for $a_{pc}$/$b_{pc}$ orientation, $a_{pc}$/$b_{pc}$ orientation that is 90$^{\circ}$ rotated in the in-plane direction compared to the previous one, and $c_{pc}$ orientation, respectively. (c) RGB composite image of Bragg filtered areas for three different type of domains. (d) RGB composite threshold image of three different type of domains.}
\end{figure}

\ch{BaSnO3}/\ch{LaInO3} heterostructures were grown on the \ch{SrTiO3} (001) substrate. Due to the high lattice mismatch of -5.5 \%, the \ch{BaSnO3} film grows relaxed on the \ch{SrTiO3} substrate. Therefore, we consider \ch{BaSnO3} as a pseudosubstrate with the bulk value of lattice parameter. Fig. \ref{fig:Fig2} (a) shows typical results of a high-resolution TEM cross-section analysis of the \ch{BaSnO3}/\ch{LaInO3} interface. The \ch{BaSnO3} film/pseudosubstrate is seen along the [010] projection. It is single crystalline and grows epitaxially along the [001] surface normal of the \ch{SrTiO3} substrate. The \ch{LaInO3} film on top of the \ch{BaSnO3} pseudosubstrate surface is characterized by three types of coherently grown 5 - 40 nm wide domains which can be distinguished by the respective image patterns (Fig. \ref{fig:Fig2} (a)). In the first domain from the left (I), the $c_{pc}$ orientation lays along the [010] projection of the \ch{BaSnO3} substrate, and can be observed along the viewing direction. The image simulation highlighted by the white frame is shown as an inset and reproduces the experimental pattern of the upper part of the domain. The center and the right domain (II and III) show a stripe-like pattern with a characteristic periodicity where each fourth plane appears at lower intensity. For the chosen imaging conditions (defocus $\Delta f = +8$ nm, thickness $t = 22.9$ nm), the symmetry of the pattern reflects the symmetry of the projected unit cell. The observed periodicity corresponds to the size of the unit cell along $a_{pc}$/$b_{pc}$, and the image simulations reproduce the experimental patterns well. In the domain III, $c_{pc}$ planes are parallel to the [001] surface normal of \ch{BaSnO3}, while in the domain II, $c_{pc}$ planes lay parallel to (100) planes of \ch{BaSnO3}. Since the $a_{pc}$ and $b_{pc}$ orientations exhibit the same atomic pattern, they cannot be distinguished from the HRTEM image pattern. They will therefore be referred to as $a_{pc}$/$b_{pc}$ in the further text. Magnified simulated patterns, overlayed by ball and stick models of the atomic structure, are shown in Figs. \ref{fig:Fig2} (b), (c), and (d). Fast Fourier transformed images of $c_{pc}$, $a_{pc}$/$b_{pc}$ which is rotated by 90$^{\circ}$ clockwise/counterclockwise, and $a_{pc}$/$b_{pc}$ are shown in Figs. \ref{fig:Fig2} (e), (f), and (g), respectively.

To get statistically significant results on the distribution of the different domains, we study the samples in plane-view TEM. Fig. \ref{fig:Fig3} (a) shows a typical image under multibeam conditions along the [001] surface normal of \ch{BaSnO3}.  Similar as in the cross-section, we assign the domains according to the orientation of the stripe-like patterns in the HRTEM images, which can be assigned to the $a_{pc}$/$b_{pc}$ planes in orthorhombic \ch{LaInO3}.  To ease the analysis, we perform Bragg filtering. We select reflections in the fast Fourier transformed image that are specific for each of the three different type of domains (Fig. \ref{fig:Fig3} (b)), and separately apply a mask on each set of specific reflexes. We then perform an inverse Fourier transformation for the specific reflection. By doing so, each of the filtered images shows only domains of the same chosen orientation. A composite Bragg filtered RGB image shown in Fig. \ref{fig:Fig3} (c) exhibits all three type of domains. Converting filtered images further to binary images, the occupancy percentage of each domain type can be calculated. Processed binary images are additionally combined together into the RGB image where red, green, and blue colors represent different domain orientations, $a_{pc}$/$b_{pc}$, $a_{pc}$/$b_{pc}$ which is rotated by 90$^{\circ}$ in the in-plane direction, and $c_{pc}$ orientation, respectively (Fig. \ref{fig:Fig3} (d)).
For quantification of the domain distribution, a total sample area of 0.4 $\mu$m\textsuperscript{2} containing approximately 1000 domains is investigated. The investigation reveals that \ch{LaInO3} domains with $c_{pc}$ orientation (blue) parallel to the \ch{BaSnO3} (001) surface are occupying 44.4 ± 5\% of the \ch{LaInO3} film, while $a_{pc}$/$b_{pc}$-oriented domains (red and green) cover together 55.6 ± 7\%. 

\subsection{\label{sec:domains}Interface}

\begin{figure}
\includegraphics[width=0.47\textwidth]{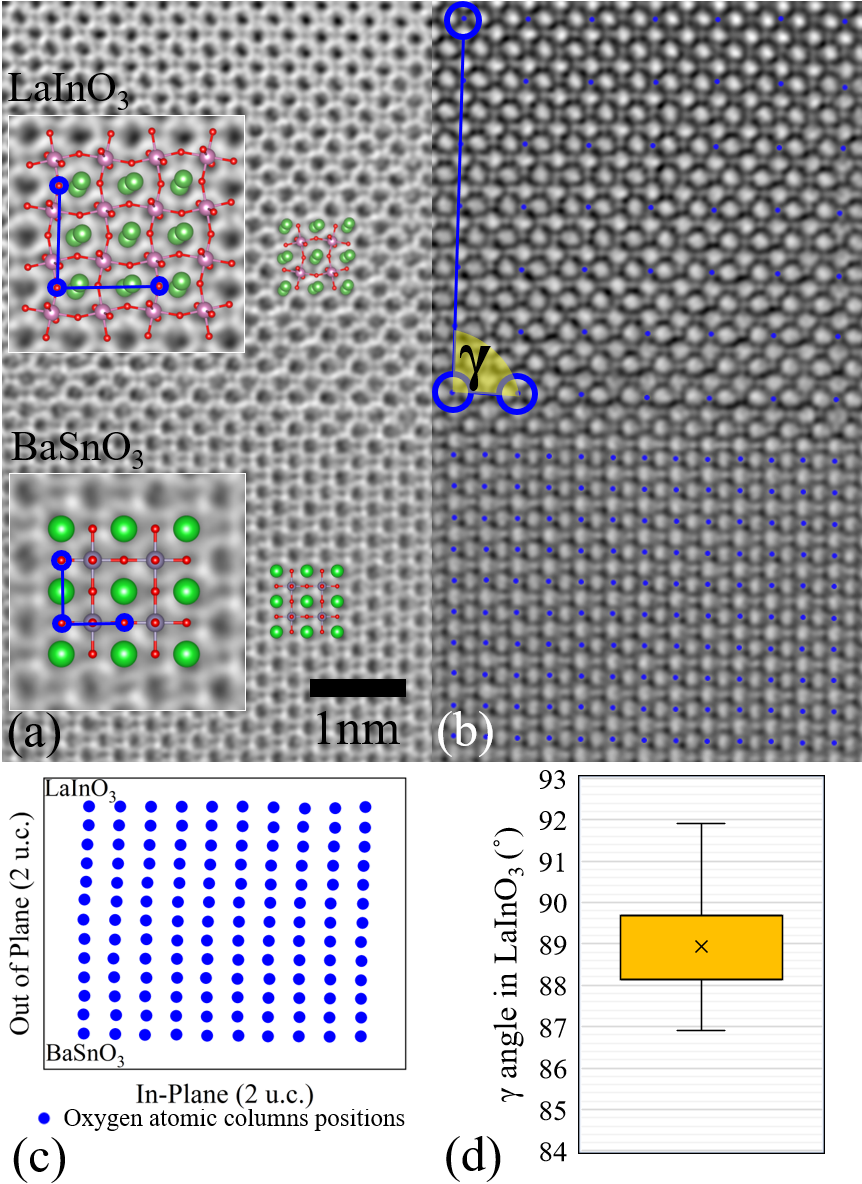}
\caption{\label{fig:Fig4}(a) Amplitude of the exit wave obtained from the exit wave reconstruction of the \ch{BaSnO3}/\ch{LaInO3} interface with insets showing the atomic structures of \ch{BaSnO3} and \ch{LaInO3}. Dark dots correspond to atoms. (b) Inversed contrast of the amplitude, used for oxygen atomic columns mapping. Oxygen positions used for $\gamma$ angle measurement are marked with blue dots, and extracted in (c). (d) Graphic representation of the $\gamma$ angle value in \ch{LaInO3}.}
\end{figure}

To get an insight into the interface structure, we study a \ch{LaInO3} domain with the $a_{pc}$/$b_{pc}$ surface normal along the $c_{pc}$ viewing direction. We focus here on the octahedral tilt, and perform exit wave reconstruction from the defocus series. Fig. \ref{fig:Fig4} (a) shows the amplitude of the exit wave reconstructed from a focal series, while Fig. \ref{fig:Fig4} (b) shows the inverse image, which is used for analysis by peak finding. The atomic structure of \ch{BaSnO3} and \ch{LaInO3} is shown as an overlay to the images. Based on the bulk properties, we expect $\gamma$ angles of 90$^{\circ}$ for \ch{BaSnO3} and of 87.6$^{\circ}$ for \ch{LaInO3}. To measure the $\gamma$ angle of \ch{LaInO3}, we perform peak finding of only equivalent oxygen atoms, corresponding to the distance of 2 pseudocubic unit cells (see Fig. \ref{fig:Fig4} (c)), and then we measure the average $\gamma$ angle across the interface as marked in Fig. \ref{fig:Fig4} (b). The result of our analysis is shown in Fig. \ref{fig:Fig4} (d). We find an $\gamma$ angle of 88.9$^{\circ}$ ± 1$^{\circ}$ in the $\text{a}_\text{pc}$-oriented \ch{LaInO3}, which is closer to 90$^{\circ}$ than to the expected equilibrium value of \ch{LaInO3} of 87.6$^{\circ}$. As a reference for our measurement, we use the 90$^{\circ}$ angle in the cubic \ch{BaSnO3} pseudosubstrate.

\subsection{\label{sec:InSitu} In-Situ Heating Experiment}

\begin{figure}
\includegraphics[width=0.47\textwidth]{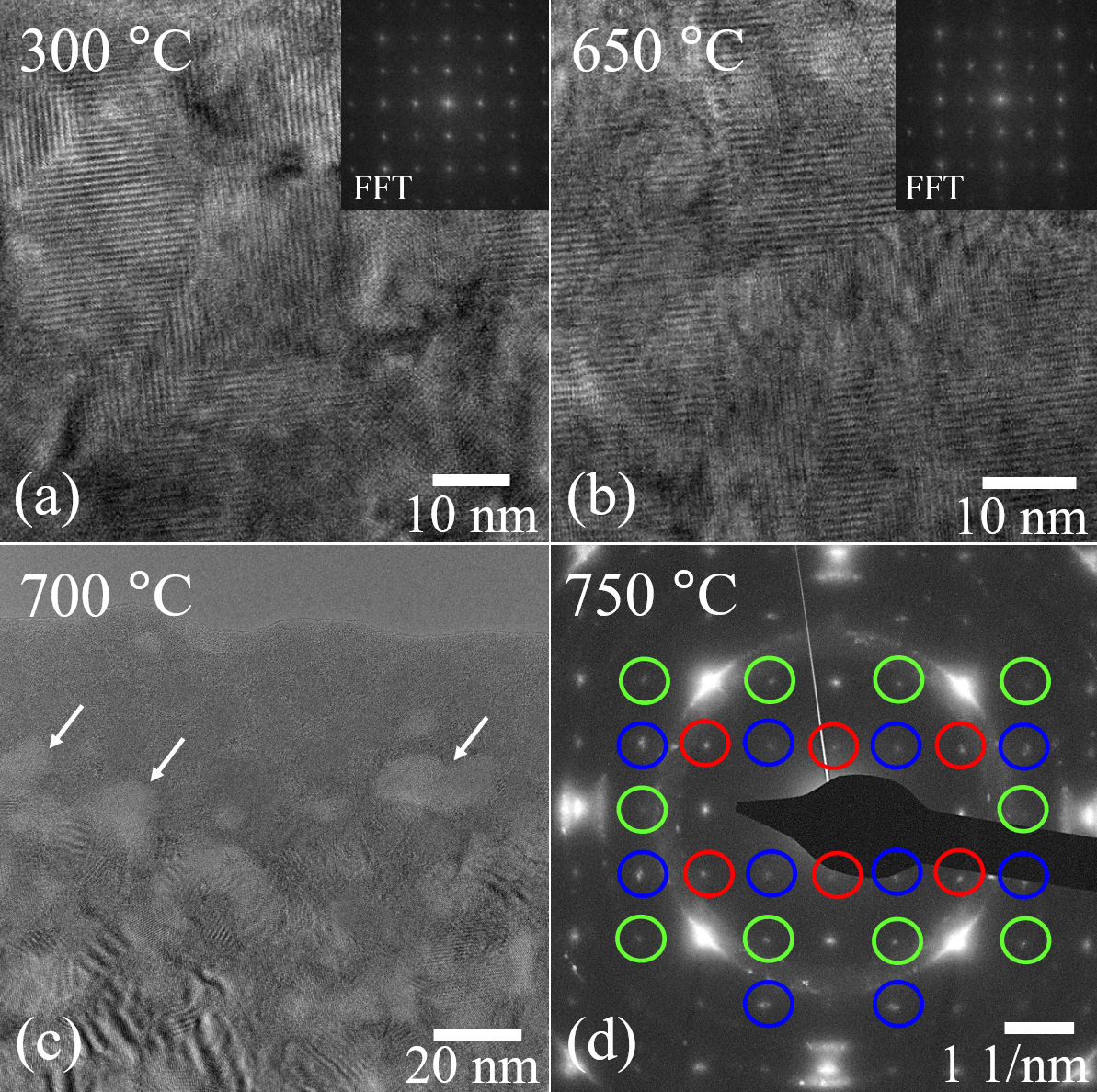}
\caption{\label{fig:Fig5}HRTEM images of the \ch{LaInO3} film heated at (a) 300 $^{\circ}$C, (b) 650 $^{\circ}$C, and (c) 700 $^{\circ}$C (white arrows are pointing at the \ch{LaInO3} film decomposition). Corresponding fast Fourier transformed images (FFT) are overlayed in (a) and (b). (d) Diffraction pattern of the \ch{LaInO3} film at 750 $^{\circ}$C, where red, green, and blue correspond to reflexes specific only for $a_{pc}$/$b_{pc}$, $a_{pc}$/$b_{pc}$ orientation that is 90$^{\circ}$ rotated in the in-plane direction, and $c_{pc}$ orientation, respectively.}
\end{figure}

To examine if \ch{LaInO3} undergoes the phase transitions between the temperature at which the film is grown and room temperature, and to check if differently oriented domains are formed immediately after the \ch{LaInO3} film growth starts or during cooling the film down, we perform an in-situ heating experiment in the temperature range required for the growth of our heterostructures. Results are shown in Fig. \ref{fig:Fig5}. Due to the thermal bending of the heating membrane, it was not possible to record HRTEM images at a fixed sample position. When analyzing the HRTEM images (Figs. \ref{fig:Fig5} (a) and (b)), features like horizontal and vertical “stripes”, that are specific for orthorhombic domains, can be recognized. Fast Fourier transformed images in Figs. \ref{fig:Fig5} (a) and (b) show that all specific reflexes for \ch{LaInO3} are present, and that the domains stay unchanged up to 650 $^{\circ}$C.  At 700 $^{\circ}$C, due to the lack of the oxygen in the in-situ experiment (which was performed in vacuum), the \ch{LaInO3} film decomposes, and formation of holes starts. The holes in the \ch{LaInO3} film are indicated by white arrows in Fig. \ref{fig:Fig5} (c). Furthermore, the diffraction pattern of the \ch{LaInO3} film at 750 $^{\circ}$C is shown in Fig. \ref{fig:Fig5} (d). Diffraction reflexes marked with red, green, and blue are specific only for $a_{pc}$/$b_{pc}$, $a_{pc}$/$b_{pc}$ orientation that is rotated by 90$^{\circ}$ in the in-plane direction, and $c_{pc}$ orientation, respectively. Presence of all three types of reflexes suggests that these domains stay stable up to 750 $^{\circ}$C. With a more detailed inspection, the elongation of the brightest reflexes that appear for all three orientations can be observed. The elongation appears due to the twisting of domains for a few degrees in the in-plane direction. The fact that reflexes specific for the three differently oriented orthorhombic domains stay stable at the growth temperature indicates that these \ch{LaInO3} domains are most likely formed through epitaxial growth and not by phase transformation during the cool-down. 

From our experimental study of coherently grown orthorhombic \ch{LaInO3} on cubic \ch{BaSnO3} we find the main results:

\begin{itemize}
\item The epitaxial layer is formed of domains that exhibit three different \ch{LaInO3} epitaxial relationships with respect to \ch{BaSnO3}, i.e. $c_{pc}$ and $a_{pc}$/$b_{pc}$ of \ch{LaInO3} parallel to the \ch{BaSnO3} (100) surface.
\item $c_{pc}$-oriented domains and $a_{pc}$/$b_{pc}$-oriented domains appear with similar probability.
\item The $\gamma$ angle of coherently strained $a_{pc}$/$b_{pc}$-oriented \ch{LaInO3} on the (001) surface of cubic \ch{BaSnO3} is close to 90$^{\circ}$, and deviates from the equilibrium value of 87.6$^{\circ}$.
\item \ch{LaInO3} domains in the \ch{LaInO3} film are formed from the beginning of the growth process.
\end{itemize}

\section{\label{sec:Calc}Computational results and discussion}

A common approach to predict the most favorable orientation relationship in epitaxial growth in heteroepitaxy is to calculate the strain energy density for different orientations considering coherent growth. The orientation with the lowest strain energy is considered as the most favorable one. Using Hooke's law $\sigma_{ij}=C_{ijkl}\cdot\varepsilon_{kl}$, the strain energy density is given by: 

\begin{equation}
\label{eq:mismatch}
W=\dfrac{1}{2}C_{ijkl}\cdot\varepsilon_{ij}\cdot\varepsilon_{kl},
\end{equation}
where $C_{ijkl}$ is the stiffness tensor that can be represented in Voigt notation as a $6\times 6$ matrix with 36 independent components, and $\varepsilon_{ij}$ represents the strain tensor. In our calculations, we consider a plane stress condition for the thin coherent \ch{LaInO3} film. This means that the stress is induced by the lattice mismatch, and present along the two perpendicular in-plane axes of the film (x- and y-axis), while elastic relaxation along the surface normal (z-direction) requires the components of the stress tensor perpendicular to the growth surface to be zero: $\sigma_{zz}=\sigma_{xz}=\sigma_{yz}=0$. The axial strain along the two in-plane axes $\varepsilon_{xx}$ and $\varepsilon_{yy}$, as well as shearing within the interfacial plane between the x- and y-axis, $\varepsilon_{xy}$, are known. The in-plane strain $\varepsilon_{xx/yy}$ in the \ch{LaInO3} layer is equal to the lattice misfit $f$ to the \ch{BaSnO3} pseudosubstrate and is given by the equation: 

\begin{equation}
\label{eq:mismatch}
f=\dfrac{a_s}{a_{l,x/y}}-1,f=\varepsilon_{xx/yy}.
\end{equation}
Here, $a_s$ is the lattice constant of the pseudosubstrate (\ch{BaSnO3}), and $a_{l,x/y}$ is the lattice constant of the \ch{LaInO3} layer along the x- or y-axis. The unknown parameters  $\varepsilon_{zz}$,  $\varepsilon_{xz}$, and  $\varepsilon_{yz}$ are calculated using Hook’s law and the boundary conditions of the stress-free surface. The elastic constants used in the \ch{LaInO3} stiffnes tensor are calculated by DFT and shown in Table \ref{tab:elasticConstants}. 

We distinguish two possible orientations: (i) growth of the $c_{pc}$-oriented domain where $(001)_{pc}$ is parallel to the (001) surface of \ch{BaSnO3}, and (ii) Growth of the $a_{pc}/b_{pc}$-oriented domains where $(100)_{pc}/(010)_{pc}$ is parallel to the  \ch{BaSnO3} (001) surface. We start with case (i). For this case we can perform the calculation in the orthorhombic coordinate system of \ch{LaInO3} with $[100]_{ortho}/[010]_{ortho}/[001]_{ortho}$ being the x-/y-/z-axis, respectively. The coherency between the epitaxial \ch{LaInO3} film and the cubic \ch{BaSnO3} pseudosubstrate requires that $a_{pc} = b_{pc} = a_{\ch{BaSnO3}}$ and that the angle $\gamma$ between $a_{pc}$ and $b_{pc}$ becomes 90$^{\circ}$. This is obtained if the two orthorhombic in-plane axes are strained such that $[100]_{ortho}=[010]_{ortho}=\sqrt{2}\cdot a_{\ch{BaSnO3}}$. No in-plane shearing between the x- and y-axis is required, i.e. $\varepsilon_{xy}=0$.

In case (ii), to calculate the strain energy density of $a_{pc}$- and $b_{pc}$-oriented films, where $(100)_{pc}$ and $(010)_{pc}$ are parallel to the \ch{BaSnO3} (001) surface, we have to rotate the coordinate system from the orthorhombic system to the pseudocubic so that the x-axis equals either $b_{pc}$ or $a_{pc}$, respectively, and $c_{pc}$ ($=c_{ortho}$) is the y-axis. Note that for case (ii) the surface normal, the z-axis, does not exactly coincide with the third pseudocubic axis $a_{pc}$ or $b_{pc}$ because the angle between $a_{pc}$ and $b_{pc}$, $\gamma$, equals 87.6$^{\circ}$ (Fig. \ref{fig:Fig1} (f)). The orthorhombic coordinate system is right hand rotated by +43.81$^{\circ}$  and -46.17$^{\circ}$ around the $c_{pc}$ axis in order to obtain $a_{pc}$- and $b_{pc}$-oriented films, respectively. The stiffness tensor for the rotated coordinate system is obtained by solving the transformation equation for a $4^{th}$ rank tensor. Summarized strain values ($\varepsilon$) for all cases are shown in Table \ref{tab:strain}. Results of the strain energy density for each of the three pseudocubic growth directions are presented in Table \ref{tab:strainenergy}.

\begin{table}[t]
\caption{\label{tab:strain}Summarized strain values ($\varepsilon$) for $\text{c}_\text{pc}$, $\text{a}_\text{pc}$ and $\text{b}_\text{pc}$ \ch{LaInO3} orientations parallel to the  \ch{BaSnO3} (001) surface.}
\centering
\footnotesize
\begin{tabular}{c c c c c c c}
\hline \hline
 \ch{LaInO3} & $\varepsilon_{xx}$ & $\varepsilon_{yy}$ & $\varepsilon_{zz}$ & $\varepsilon_{yz}$ & $\varepsilon_{zx}$ & $\varepsilon_{xy}$\\ \hline
$c_{pc}$ & -0.019 & 0.022 & - & 0 & 0 & 0 \\ 
$a_{pc}$ & $8.9\cdot10^{-4}$ & $3.4\cdot10^{-3}$ & -0.002 & $1.6\cdot10^{-4}$ & 0 & 0\\
$b_{pc}$ & $8.9\cdot10^{-4}$ & $3.4\cdot10^{-3}$ & -0.002 & -$1.6\cdot10^{-4}$ & 0 & 0\\ \hline\hline
\end{tabular}
\end{table}

\begin{table}[t]
\caption{\label{tab:strainenergy}Calculated strain energy density for all three growing directions.}
\centering
\footnotesize
\begin{tabular}{c c c c}
\hline \hline
 Orientation parallel to the (001)& $a_{pc}$ & $b_{pc}$ & $c_{pc}$\\ 
\ch{BaSnO3} surface & & &\\ \hline
 Strain energy density (GPa) & 0.0011 & 0.0011 & 0.0480 \\ \hline
\hline
\end{tabular}
\end{table}

Considering that the area occupancy of the differently oriented domains scales with the strain energy density, we found an obvious discrepancy to the data obtained from our TEM analysis. While the calculated strain energy density of  $a_{pc}$/$b_{pc}$ domains (shown in Tab. \ref{tab:strainenergy}) is by an order of magnitude lower than the one of the  $c_{pc}$ domains, both orientations occur with similar probability in the TEM analysis. To address this question, we compute the formation energy of bulk and strained orthorhombic \ch{LaInO3} for different orientations using DFT. We focus on the influence of the experimentally observed deviation of the angle $\gamma$ from the bulk value in the $a_{pc}$- and $b_{pc}$-oriented layers. Since  $a_{pc}$- and $b_{pc}$-oriented \ch{LaInO3} are symmetry equivalent, we restrict ourselves to $a_{pc}$- and $c_{pc}$-oriented \ch{LaInO3}. The formation energy is calculated for a unit cell of \ch{LaInO3} with periodic boundary conditions. For $a_{pc}$-oriented \ch{LaInO3}, the primitive unit cell is formed by $2 \times 2 \times 2$ pseudocubic unit cells (see Fig. \ref{fig:Fig1} (f)). For the $c_{pc}$-oriented \ch{LaInO3}, the primitive cell is shown in Fig. \ref{fig:Fig1} (a). The in-plane parameters are fixed to the value of the \ch{BaSnO3} pseudosubstrate, while the atomic positions and the out-of-plane parameters are relaxed. This means that the calculation mimics rather a thick film, but still a coherently strained one. For the $a_{pc}$ orientation along the growth direction, we calculate two distinct cases with (i) $\gamma$ fixed to 87.6$^{\circ}$ like in the bulk crystal, and with (ii) $\gamma$ fixed to 90$^{\circ}$, which is closer to the experimentally observed angle $\gamma$ in our epitaxial layers. The formation energy, $E_f$, is computed as:
\begin{equation}
\label{eq:eform}
E_f=E^{\ch{LaInO3}}_{tot}-\sum _\text{\textit{elements}}E_{tot}^{bulk},
\end{equation} 
where $E^{\ch{LaInO3}}_{tot}$ and $E_{tot}^{bulk}$ are the total energies of orthorhombic \ch{LaInO3} in the crystalline state and of its constituent elements (La, In, O) in their reference states, respectively. 

\begin{table}[t]
\caption{\label{tab:eformbulk}Formation energies ($E_f$) per formula unit (f.u.) of $a_{pc}$- and $c_{pc}$-oriented \ch{LaInO3}.}
\centering
\footnotesize
\begin{tabular}{c c@{\hskip 15pt} c c c}
\hline \hline
\multirow{2}{*}{\ch{LaInO3}} & \multicolumn{3}{c}{Lattice parameters (\AA)} & \\
 & a & b & c & $E_f (eV/f.u.)$\\ \hline
$a_{pc}$ out-of-plane ($\gamma = 87.6^{\circ}$) & 8.23 & 8.23 & 8.20 & -12.86\\
$a_{pc}$ out-of-plane ($\gamma = 90^{\circ}$)& 8.23 & 8.23 & 8.20 & -12.84\\ 
$c_{pc}$ out-of-plane & 5.82 & 5.82 & 8.20 & -12.84 \\ \hline
\hline
\end{tabular}
\end{table}

Table \ref{tab:eformbulk} summarizes the formation energies per formula unit of bulk \ch{LaInO3} for the three different cases, i.e. \ch{LaInO3} coherently strained to \ch{BaSnO3} (001) with $c_{pc}$ and $a_{pc}$ orientation. The \ch{LaInO3} film with $a_{pc}$ along the growth direction, with $\gamma$ = 87.6$^{\circ}$ (as in the bulk crystal), has the lowest formation energy indicating that it would be the most favorable structure. This result is, as expected, in agreement with that from continuum elasticity theory (see Table \ref{tab:strainenergy}), where the orientation along $a_{pc}$ is more favorable than the orientation along $c_{pc}$. Although the difference in formation energy between $a_{pc}$ ($\gamma$ = 87.6$^{\circ}$) and $c_{pc}$ is only 0.02 eV/f.u., corresponding to 0.0461 GPa, that matches very well the calculated strain energy density value of 0.0480 GPa for the $c_{pc}$-oriented domain. For $a_{pc}$-oriented \ch{LaInO3} with $\gamma$ = 90$^{\circ}$, however, the formation energy is similar to the formation energy of $c_{pc}$-oriented \ch{LaInO3}. This finding is consistent with the experimentally observed distribution of $a_{pc}$- and $c_{pc}$-oriented domains.

\begin{figure}
\includegraphics[width=0.47\textwidth]{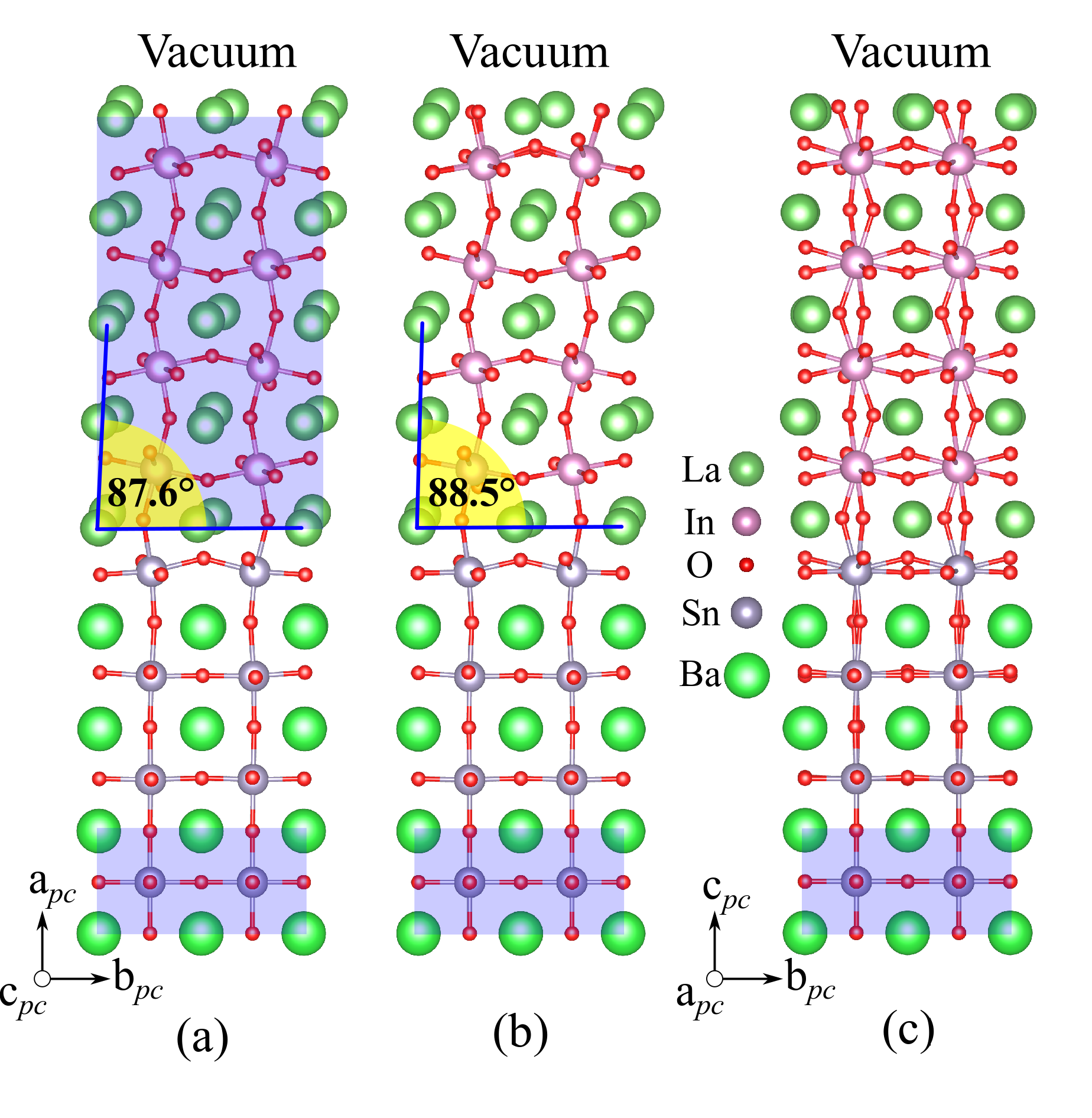}
\caption{\label{fig:Fig6}Geometries of coherently strained $a_{pc}$ ((a) and (b)), and $c_{pc}$ (c) orientation of \ch{LaInO3} on top of \ch{BaSnO3} (001) surface. (a) In \ch{LaInO3}, the blue shadowed region marks fixed geometry. Blue shadowed regions in \ch{BaSnO3} mark the first unit cell which has lattice parameter fixed to the bulk value.}
\end{figure}

In the following, we will address the influence of the \ch{BaSnO3} pseudosubstrate and the \ch{LaInO3}/\ch{BaSnO3} interface, focusing on their effect on the angle $\gamma$ in the coherently strained \ch{LaInO3} film. We compare formation energies per atom in a heterostructure where \ch{LaInO3} is coherently strained on \ch{BaSnO3}. In our calculation, a supercell consisting of a very thin film of 4 pseudocubic unit cells of \ch{LaInO3} on top of a \ch{BaSnO3} pseudosubstrate is considered. This calculation explicitly includes the interface and its effect on the domain formation. In the case shown in Fig. \ref{fig:Fig6} (a) the $a_{pc}$ orientation of the \ch{LaInO3} film is parallel to the \ch{BaSnO3} (001) surface and $\gamma$ is fixed to 87.6$^{\circ}$. Fig. \ref{fig:Fig6} (b) represents the case where \ch{LaInO3} is relaxed. For comparison, Fig. \ref{fig:Fig6} (c) shows a $c_{pc}$-oriented film  where \ch{LaInO3} is relaxed. A striking observation here is that upon relaxation, the angle $\gamma$ changes from the equilibrium value of 87.6$^{\circ}$ to 88.5$^{\circ}$. This behavior is in excellent agreement with our TEM measurements.

\begin{table}[t]
\caption{\label{tab:eforminterface}Summarized formation energies in eV/atom for coherent growth of \ch{LaInO3} on the \ch{BaSnO3} (001) surface.}
\centering
\footnotesize
\begin{tabular}{c c c c}
\hline \hline
 \ch{LaInO3} & $a_{pc}$ ($\gamma = 87.6^{\circ}$) & $a_{pc}$ ($\gamma = 88.5^{\circ}$) & $c_{pc}$\\ \hline
$E_f$ & -2.419 & -2.429 & -2.428\\ \hline
\hline
\end{tabular}
\end{table}

Table \ref{tab:eforminterface} summarizes the formation energies per atom for all three cases shown in Fig. \ref{fig:Fig6}. Comparing the formation energy of $a_{pc}$ with two different values of $\gamma$, we find a behavior opposite to that of the bulk \ch{LaInO3} formation energies reported in Table \ref{tab:eformbulk}. The $a_{pc}$ orientation with $\gamma$ angle as in the bulk is now energetically less favorable than the $a_{pc}$ orientation where $\gamma$ angle is closer to 90$^{\circ}$. The increase of the $\gamma$ angle (Fig. \ref{fig:Fig6} (b)) decreases the formation energy for the $a_{pc}$ orientation, and shifts it towards that of the $c_{pc}$ orientation. Therefore, the formation energies of relaxed $a_{pc}$ ($\gamma$ closer to 90°) and $c_{pc}$ orientation become similar, which is in excellent agreement with the domain distribution observed by TEM. This confirms that here the interface is decisive in controlling the epitaxial relationship, instead of the exclusive consideration of continuum mechanics, i.e. strain energy. 

\begin{figure}
\includegraphics[width=0.38\textwidth]{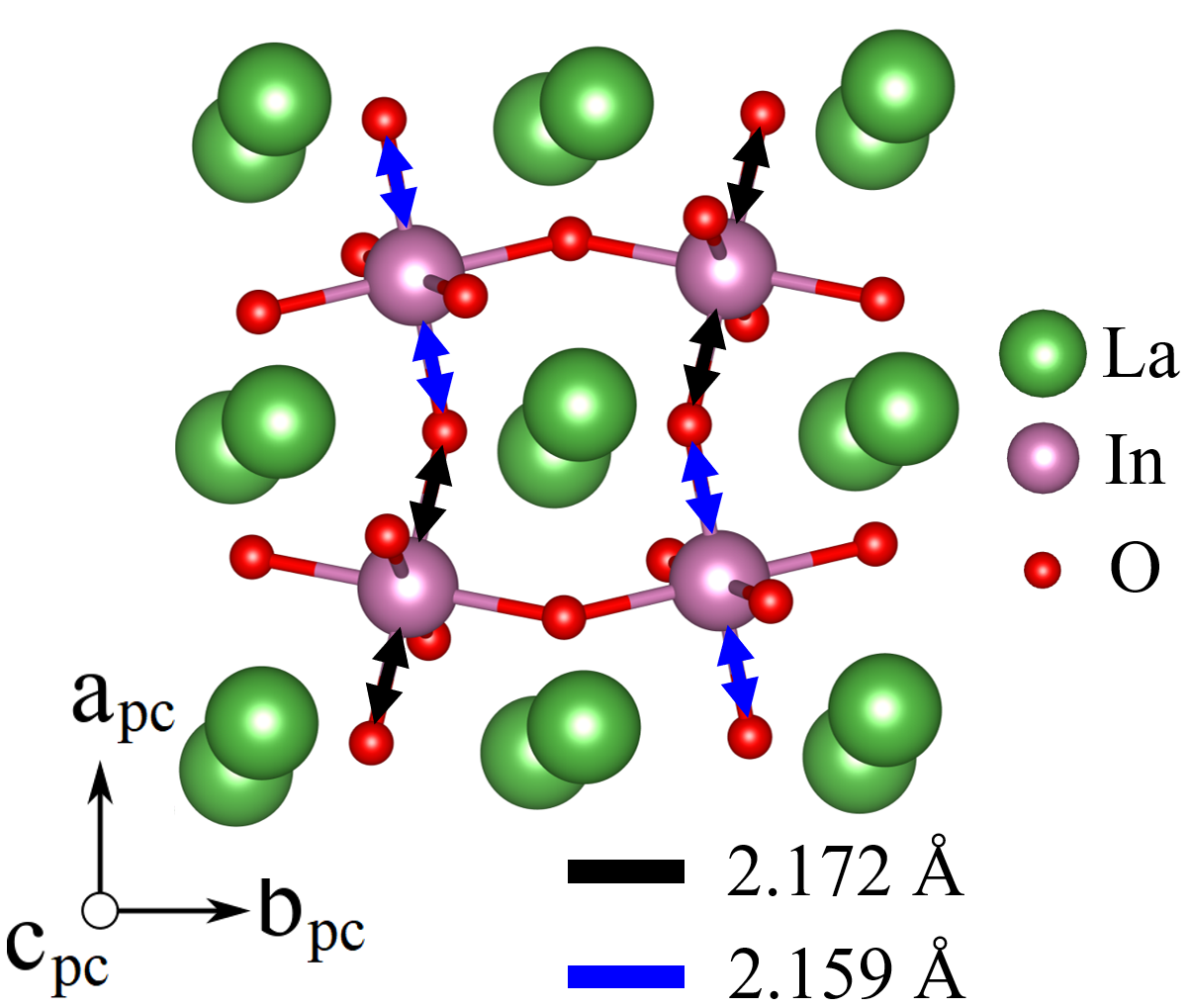}
\caption{\label{fig:Fig7}Alternation of In-O bond lenghts in bulk \ch{LaInO3}.}
\end{figure}

To get insight into the role of the interface, we have a closer look on the octahedra tilt pattern of the $a_{pc}$-oriented interface, and compare it to that of bulk \ch{LaInO3} and bulk \ch{BaSnO3}. As shown in Fig. \ref{fig:Fig7}, in bulk \ch{LaInO3}, an alternation of In-O bond lengths within the inequivalent octahedra along the $c_{pc}$ direction can be observed ($a_{pc}$ in the out of plane direction). In contrast, there is no such alternation of the Sn-O bond length in the cubic perovskite \ch{BaSnO3}. Moreover, these bonds are shorter and therefore stronger than the In-O bonds in \ch{LaInO3}. Once $a_{pc}$-oriented \ch{LaInO3} grows on top of \ch{BaSnO3} (001), the stronger Sn-O bonds force the In-O bonds within the inequivalent octahedra to become equal in length at the interface. In addition, the in-plane La shift is reduced due to the lack of oxygen octahedral tilt in the \ch{BaSnO3} substrate. Therefore, it is the local chemistry of bonding at the interface, i.e. the change in In-O bond length and the shift of the La atoms at the interface due to the \ch{BaSnO3} pseudosubstrate, that leads to an increase of the angle $\gamma$ to almost 90$^{\circ}$. This increase shifts the formation energy in the $a_{pc}$-oriented \ch{LaInO3} domains. Such balancing of formation energies has a strong impact on the distribution of domains in epitaxial films, i.e both, $a_{pc}$ and $c_{pc}$, orientations are equally distributed, although from strain energy arguments we would expect a strong preference for the $a_{pc}$ orientation.

\section{Summary and Conclusions}

In this work, we have presented a combined experimental and theoretical study of the domain structure in orthorhombic \ch{LaInO3} thin films coherently grown on a cubic \ch{BaSnO3} pseudosubstrate. We have shown that the epitaxial layer is formed of domains that correspond to the three different pseudocubic orientations of \ch{LaInO3}. In contrast to a common approach that predicts the preferred orientation relationship between film and substrate based on strain energy density calculations, we have shown that the interface chemistry of bonding may balance the energetics of the system and counteract the strain accommodation. Our finding highlights the remarkable influence of the interface chemistry on the epitaxial relationship in heterostructures of different \ch{ABO3} perovskites with different symmetries. 

\section{Acknowledgements}
This work was supported by the Leibniz Senatsausschuss Wettbewerb (SAW) project BaStet (No. K74/2017) and was performed in the framework of GraFOx, a Leibniz science campus partially funded by the Leibniz Association. We thank the European Community (Europaeischer Fonds für regionale Entwicklung-EFRE) under Grant No. 1.8/15 for partially funding this project. W. A. acknowledges the North-German Supercomputing Alliance (HLRN) for providing HPC resources that have contributed to the research results reported in this paper - project bep00078. We thank Tore Niermann, Technical University Berlin, for partial support in TEM work. Useful discussion with Tobias Schulz is gratefully acknowledged.
 
\section{References}
 \renewcommand{\section}[2]{}
\bibliography{references}

\end{document}